\documentclass[
aps,nofootinbib,showpacs,showkeys,preprint, 
]
{revtex4}

\usepackage{epsf,epsfig,subfigure,axodraw,graphicx,amsmath,amssymb}
\usepackage{color}

\begin{document}

\begin{flushright}
{\tt 
~~~~SNUTP 09-012 \quad
\\}
\end{flushright}

\title{\Large\bf
Decaying LSP in SO(10) GUT and \\ PAMELA's Cosmic Positron}

\author{\large
Bumseok Kyae\email{bkyae@phya.snu.ac.kr} } \affiliation{
Department of Physics and Astronomy and Center for Theoretical
Physics, Seoul National University, Seoul 151-747, Korea  }

\vspace{0.7cm}

\begin{abstract}

We suppose that the lightest supersymmetric particle (LSP) in the
minimal supersymmetric standard model (MSSM) is the dark matter.
The bino-like LSP can decay through the SO(10) gauge interactions,
if one right-handed (RH) neutrino ($\nu^c_1$) is lighter than the
LSP and its superpartner ($\tilde{\nu}^c_1$) develops a vacuum
expectation value (VEV), raising extremely small R-parity
violation naturally. The leptonic decay modes can be dominant, if
the VEV scale of ${\bf 16}_H$ is a few orders of magnitude lower
than the VEV of ${\bf 45}_H$ ($\approx 10^{16}$ GeV), and if a
slepton ($\tilde{e}^c_1$) is relatively lighter than squarks. The
desired decay rate of the LSP, $\Gamma_\chi\sim 10^{-26}$
sec$.^{-1}$ to explain PAMELA data can be naturally achieved,
because the gaugino mediating the LSP decay is superheavy. From
PAMELA data, the
SU(3)$_c\times$SU(2)$_L\times$SU(2)$_R\times$U(1)$_{\rm B-L}$
breaking scale (or the ${\bf 16}_H$ VEV scale) can be determined.
A global symmetry is necessary to suppress the Yukawa couplings
between one RH (s)neutrino and the MSSM fields. Even if one RH
neutrino is quite light, the seesaw mechanism providing the
extremely light three physical neutrinos and their oscillations is
still at work.

\end{abstract}

\pacs{12.10.Dm, 95.35.+d, 95.30.Cq, 12.60.Jv}

\keywords{SO(10), Dark matter decay, Cosmic positron excess}
\maketitle

\section{Introduction}

For the last three decades, many remarkable progresses in particle
physics and cosmology have been made thanks to the cooperateive
and intimate relation between the two fields. In particular, the
application of particle physics theory into dark matter (DM)
models in cosmology was very successful. Because of the correct
order of magnitude of the cross section, thermally produced weakly
interacting massive particles (WIMPs) have been long believed to
be DM candidates \cite{WIMP}. So far the lightest supersymmetric
particle (LSP), which is a well-motivated particle originated from
the promising particle physics model, i.e. the minimal
supersymmetric standard model (MSSM), has attracted much
attentions as an excellent example of WIMP.

Recently, PAMELA \cite{PAMELA}, ATIC \cite{ATIC}, H.E.S.S.
\cite{HESS}, and the Fermi-LAT collaborations \cite{Fermi-LAT}
reported the very challenging observations of positron excesses in
cosmic ray above 30 GeV upto the TeV scale.
In particular, PAMELA observed a positron fraction
[$e^+/(e^++e^-)$] exceeding the theoretical expectation
\cite{Moskalenko:1997gh} above 30 GeV upto 100 GeV. However, the
anti-proton/proton flux ratio was quite consistent with the
theoretical calculation. The ATIC, H.E.S.S., and Fermi-LAT's
observations exhibit excesses of $(e^++e^-)$ flux in cosmic ray
from 100 GeV to 1 TeV.\footnote{H.E.S.S. measured Cherenkov
radiations by cosmic electrons and positrons above 600 GeV energy
scale.} They would result from the positron flux that keeps rising
upto 1 TeV.

Apparently the above observational results are very hard to be
interpreted in view of the conventional MSSM cold dark matter
scenario: explaining the excess positrons with annihilations of
Majorana fermions such as the LSP needs a too huge boost factor.
Moreover, ATIC, H.E.S.S., and Fermi-LAT's observations seem to
require a TeV scale DM, if they are caused indeed by DM
annihilation or decay. Introduction of a TeV scale LSP, however,
would spoil the motivation of introducing supersymmetry (SUSY) to
resolve the gauge hierarchy problem in particle physics. In
addition, TeV scale DM seems to be disfavored by the gamma ray
data \cite{HESS}, if the excess positron flux is due to DM
annihilations \cite{GammaConst}. On the other hand, the DM decay
scenario is relatively free from the gamma ray constraint
\cite{strumia}.

In the DM decay scenario, however, there are some serious hurdles
to overcome: one is to naturally obtain the extremely small decay
rate of the DM ($\Gamma_{\rm DM}\sim 10^{-26}$ sec.$^{-1}$), and
the other is to naturally explain the relic density of the DM in
the Universe. The first hurdle could be somehow resolved by
introducing an extra symmetry, an extra DM component with a TeV
scale mass, and grand unified theory (GUT) scale superheavy
particles, which mediate DM decay into the SM charged leptons (and
the LSP) \cite{DMdecayModel}. The fact that the GUT scale
particles are involved in the DM decay might be an important hint
supporting GUT \cite{flippedSU5,Dimopoulos}. However, since the
interaction between the new DM and the SM charged lepton are made
extremely weak by introducing superheavy particles mediating the
DM decay, non-thermal production of the DM with a carefully tuned
reheating temperature should be necessarily assumed. One way to
avoid it is to consider SUSY models with two DM components
\cite{DMdecayModel,flippedSU5}. In these models, the decay of the
small amount of the meta-stable heavier DM component ($X$), which
is assumed to be non-thermally produced, accounts for the cosmic
positron excess, and the thermally produced lighter DM component
LSP ($\chi$), which is absolutely stable and regarded as the
dominant DM [${\cal O}(10^{-10})<n_{X}/n_{\chi}$], explains the
relic density of the Universe.\footnote{The low energy field
spectrum in the models of Ref.~\cite{DMdecayModel} is the same as
that of the MSSM except for the neutral singlet extra DM
component. Moreover, the models in \cite{DMdecayModel} can be
embedded in the flipped SU(5) GUT and string models
\cite{flippedSU5,stringMSSM}.}

In this paper, we suppose that the conventional bino-like LSP is
the main component of the DM.  Since the ``bino'' is a WIMP,
thermally produced binos could explain well the relic density of
the Universe. The bino-like LSP with a mass of about 300 -- 400
GeV could also explain PAMELA data, if it decays to $e^\pm$ and a
neutral fermion with an extremely small decay rate of order
$10^{-26}$ sec.$^{-1}$ \cite{decayPAMELA}. The $(e^++e^-)$ excess
observed by Fermi-LAT could be explained by astrophysical sources
such as nearby pulsars \cite{kane} (and/or with the sub-dominant
extra TeV scale DM component
\cite{DMdecayModel}).\footnote{Alternatively, one could assume a
bino mass of 3.5 TeV in order to account for both PAMELA and
Fermi-LAT with LSP decay \cite{decayPAMELA}.  In this case,
however, the soft SUSY breaking scale should be higher than 3.5
TeV.} In fact, pulsars can explain both the PAMELA and Fermi-LAT's
data in a suitable parameter range \cite{profumo}. However, this
does not imply that DM in addition to pulsars can not be the
source of the galactic positrons \cite{kane}. In fact, we don't
know yet a complete pulsar model, in which all the free parameters
would be fixed by the fundamental physical constants.
%

To achieve the needed extremely small decay rate of the bino-like
LSP $\chi$, we need extremely small R-parity violation. We will
assume that the R-parity is broken by a non-zero vacuum
expectation value (VEV) of a right-handed (RH) sneutrino
($\langle\tilde{\nu}_1\rangle\neq 0$). Since it doesn't carry any
standard model (SM) quantum number, it does not interact with the
MSSM fields at all, if its Yukawa interactions with them are
forbidden by a symmetry and gravity interaction is ignored. We
will explore the possibility that the extremely small DM decay
rate results from the {\it gauge interaction} by exchange of the
superheavy gauge bosons and gauginos present in the SO(10) SUSY
GUT. We will not introduce a new DM component, and will attempt to
explain the PAMELA's observation within the framework of the
already existing particle physics model.

\section{SO(10) GUT}

One of the appealing GUTs is the SO(10) GUT \cite{SO(10)}. It
unifies all the three SM gauge forces within the SO(10) gauge
interaction. One of the nice features of SO(10) is that it
predicts the existence of the RH neutrinos [or the SU(2)$_L$
singlet neutrinos], since a RH neutrino is contained in a single
spinorial representation ${\bf 16}$ of SO(10), together with one
family of the SM fermions. The RH neutrinos provide a very nice
explanation of the observed neutrino oscillations through the
seesaw mechanism \cite{seesaw} and also of the baryon asymmetry in
the Universe through leptogenesis \cite{leptogenesis}.

\subsection{Superheavy fields in SO(10)}

SO(10) GUT models contain many superheavy particles. They might be
utilized to get the required DM decay rate of $10^{-26}$
sec.$^{-1}$  Most of all, the gauge bosons and gauginos
corresponding to the coset SO(10)/SM have masses around the GUT
scale. In this paper, we are particularly interested in them as
the mediators of DM decay.

The superfields in the Higgs sector needed for breaking SO(10) to
the SM are also superheavy. Particularly, an adjoint Higgs ${\bf
45}_H$ (or ${\bf 210}_H$) and spinorial Higgs ${\bf 16}_H$ and
$\overline{\bf 16}_H$ (or ${\bf 126}_H$ and $\overline{\bf
126}_H$) can be employed to achieve the SM gauge group from
SO(10). The vector representation ${\bf 10}_h$, which includes the
two MSSM Higgs doublets, containes also the superheavy Higgs
triplets $D_H$ and $D^c_H$. Their masses can be obtained through a
proper doublet/triplet splitting mechanism. One way is to
introduce the coupling ${\bf 10}_h{\bf 45}_H {\bf 10}_h$, assuming
the scalar component of ${\bf 45}_H$ develops a VEV along the
SU(3)$_c\times$SU(2)$_L\times$SU(2)$_R\times$U(1)$_{B-L}$ ($\equiv
$ LR) direction \cite{DWmech}.\footnote{When SO(10) is broken by
${\bf 16}_H$, $\overline{\bf 16}_H$, and ${\bf 45}_H$, and the
doublets/triplets in ${\bf 10}_h$ are split by the coupling ${\bf
10}_h\langle {\bf 45}_H\rangle {\bf 10}_h$, the pseudo-goldstones
included in the Higgs would not become easily superheavy. Then the
Higgs sector needs to be extended by introducing more superfields
and specific interactions \cite{minimalSO(10)}.} In this paper, we
will thus identify the triplet Higgs mass scale with the VEV of
${\bf 45}_H$.

How many and what kind of Higgs fields are needed to get the SM
gauge group are quite model-dependent. Their masses would be close
to the GUT scale, but they are not exactly the same as each other.
Even in one Higgs multiplet, its component fields might have
various mass spectra after symmetry breaking. Except for ${\bf
10}_h\langle {\bf 45}_H\rangle {\bf 10}_h$, they interact with the
MSSM fields only through non-renormalizable Yukawa couplings due
to their GUT scale VEVs.  Such couplings can be utilized to get
the realistic SM fermion masses.
One might think that SO(10)-breaking superheavy fields also
contribute to the mediation of DM decay through such
non-renormalizable couplings with the MSSM fields. However, the
extra suppression factor $(1/M_P)^n$ ($n=1,2,3,\cdots$) makes
their contributions negligible compared to those of the superheavy
gauge fields and gauginos via the renormalizable gauge
interactions, which will be discussed later.

The SO(10)-breaking sector could include heavy fields, which do
not develop GUT scale VEVs. They are introduced in order to
decouple unwanted fields in the SO(10)-breaking Higgs sector,
which are absent in the MSSM, from low energy physics in
non-minimal SO(10) models. Since their couplings to the MSSM
fields are not essential and their masses would be heavier than
the mediators leading to DM decay, we can assume that all the
interactions between such SO(10)-breaking sector fields and the
MSSM fields are weak enough, if they are present.

Thus, as far as the DM decay is concerned, the gauge interactions
through the superheavy gauge fields and gauginos can be dominant
over Yukawa interactions. They would give more predictable
results, regardless of what specific SO(10) models are adopted. We
will focus on the DM decay predominantly through the superheavy
gauge fields or gauginos.

\subsection{SU(5) vs. SU(2)$_{R}$ scale}

In terms of the SM's quantum numbers, the SO(10) generator ($={\bf
45}_G$) is split into the SM gauge group's generators plus $\{{\bf
(1,1)}_{-1}, {\bf (1,1)}_{1}\}$, ${\bf (1,1)}_0$, $\{{\bf
(3,2)}_{-5/6},{\bf (\overline{3},2)}_{5/6}\}$, and $\{{\bf
(3,2)}_{1/6},{\bf (\overline{3},2)}_{-1/6};~{\bf (3,1)}_{2/3},{\bf
(\overline{3},1)}_{-2/3}\}$.  We will simply write them as
\begin{eqnarray}
 \{E,E^c\}~,~~N~,~~\{Q',Q^{'c}\}~,~~
\{Q,Q^c~;~U,U^c\}~,
\end{eqnarray}
respectively.  By the VEV of the adjoint Higgs $\langle {\bf
45}_H\rangle$, the SO(10) gauge symmetry may break to LR. Through
this process, the gauge boson and the gauginos carrying the
quantum numbers of $\{Q',Q^{'c}\}$ and $\{Q,Q^{c}~;~U,U^c\}$
achieve heavy masses proportional to $\langle {\bf 45}_H\rangle$.
The $\{E,E^c\}$ and a linear combination of the SM hypercharge
generator and $N$ ($\equiv N_R$) composes the SU(2)$_R$
generators. The other combination orthogonal to it corresponds to
the U(1)$_{B-L}$ generator ($\equiv N_{BL}$). They don't get
masses from $\langle {\bf 45}_H\rangle$.

On the other hand, the VEVs of the Higgs in the spinorial
representations $\langle {\bf 16}_H\rangle$, $\langle
\overline{\bf 16}_H\rangle$ breaks SO(10) down to SU(5). This
process generates the heavy masses proportional to $\langle {\bf
16}_H\rangle$ ($=\langle \overline{\bf 16}_H\rangle$ in the SUSY
limit) for the gauge bosons and their superpartners of
$\{E,E^c\}$, $N$, and $\{Q,Q^{c}~;~U,U^c\}$. The SO(10) gauge
bosons associated with $\{Q',Q^{'c}\}$ correspond to the so-called
``$X$ and $Y$'' gauge bosons in SU(5). Hence, $\langle {\bf
45}_H\rangle$ and $\langle {\bf 16}_H\rangle$ determine the SU(5)
and LR breaking scales, respectively.

Alternatively, one can employ the large representations, ${\bf
126}_H$, $\overline{\bf 126}_H$, and ${\bf 210}_H$, instead of
${\bf 16}_H$, $\overline{\bf 16}_H$, and ${\bf 45}_H$
\cite{LargeRep}. ${\bf 126}_H$ and $\overline{\bf 126}_H$ break
SO(10) to SU(5), while ${\bf 210}_H$ breaks SO(10) to
SU(4)$_c\times$SU(2)$_L\times$SU(2)$_R$. In our discussion
throughout this paper, ${\bf 16}_H$ ($\overline{\bf 16}_H$) and
${\bf 45}_H$ can be replaced by ${\bf 126}_H$ ($\overline{\bf
126}_H$) and ${\bf 210}_H$, respectively.

Non-zero VEVs of both $\langle {\bf 45}_H\rangle$ and $\langle
{\bf 16}_H\rangle$ eventually give the SM gauge symmetry at low
energies. If $\langle {\bf 45}_H\rangle>\langle {\bf
16}_H\rangle$, SO(10) is broken first to LR at a higher energy
scale and further broken to the SM gauge group at lower energy
scales.  On the other hand, if $\langle {\bf 45}_H\rangle<\langle
{\bf 16}_H\rangle$, SO(10) is broken first to SU(5) at a higher
energy scale and then eventually to the SM gauge group at lower
energy scales.
%
%
While the SU(5) breaking scale by $\langle {\bf 45}_H\rangle$
could be inferred from the renormalization group (RG) running
effects of the three MSSM gauge couplings to be of $3\times
10^{16}$ GeV, the LR (or equivalently ${\rm B-L}$) breaking scale
by $\langle {\bf 16}_H\rangle$ may not be pinned down in
principle: from the seesaw mechanism for the extremely light
neutrinos, the LR breaking scale is just roughly estimated to be
around $10^{16}$ GeV. However, one should note that when the
physical neutrino mass scale is theoretically estimated through
the seesaw mechanism, the unknown Yukawa couplings associated with
the RH neutrinos are involved. Moreover, the absolute neutrino
masses can not be determined from the solar and atmospheric
neutrino oscillations.

Thus, if $\langle {\bf 45}_H\rangle>\langle {\bf 16}_H\rangle
=\langle \overline{\bf 16}_H\rangle \neq 0$ ($\langle {\bf
16}_H\rangle =\langle \overline{\bf 16}_H\rangle > \langle {\bf
45}_H\rangle \neq 0$), the gauge bosons and gauginos of
$\{Q',Q^{'c}\}$ achieve heavier (lighter) masses than those of
$\{E,E^c\}$ and $N$. The masses of the gauge sectors for
$\{Q,Q^{c}~;~U,U^c\}$ would be given dominantly by the heavier
masses in any cases, since both $\langle {\bf 45}_H\rangle$ and
$\{\langle {\bf 16}_H\rangle, \langle \overline{\bf
16}_H\rangle\}$ contribute to their masses. Accordingly, the
comparison of e.g. the gaugino masses of $\{Q',Q^{'c}\}$ and
$\{E,E^c\}$ ($\equiv M_{Q'}, M_{E}$, respectively) could determine
the hierarchy between $\langle {\bf 45}_H\rangle$ and $\langle
{\bf 16}_H\rangle$, and so the SO(10) breaking pattern too.

\section{LSP decay in SO(10)}

\begin{figure}[t]
\begin{center}
\begin{picture}(400,115)(-10,0)
\ArrowLine(45,90)(-9,90)
\Photon(-9,90)(45,90){2.5}{4}\ArrowLine(45,90)(85,115)
\DashArrowLine(65,50)(45,90){3}
\ArrowLine(100,50)(65,50)\ArrowLine(100,50)(135,50)
\Line(65,50.5)(135,50.5)\Line(65,49.5)(135,49.5)
\Photon(65,50)(135,50){3}{5} \ArrowLine(40,10)(65,50)
\DashArrowLine(135,50)(145,65){3} \ArrowLine(162,10)(135,50)
\Text(102,50)[]{\LARGE $\times$} \Text(102,60)[]{$m_{3/2}$}
\Text(20,102)[]{$\chi$} \Text(150,75)[]{$\langle
\tilde{\nu}^{c*}_1\rangle$} \Text(65,113)[]{$e^c_1$}
\Text(42,67)[]{$\tilde{e}^{c*}_1$} \Text(42,30)[]{$\nu^c_1$}
\Text(162,30)[]{$e^c_1$}
\Text(85,35)[]{$\tilde{E}^c$}\Text(115,35)[]{$\tilde{E}$}
\Text(53,50)[]{$g_{10}$} \Text(152,50)[]{$g_{10}$}
%
\DashArrowLine(255,50)(230,90){3}
\DashArrowLine(365,50)(390,90){3}
%
\ArrowLine(310,50)(255,50) \ArrowLine(310,50)(365,50)
\Line(365,50.5)(255,50.5) \Line(365,49.5)(255,49.5)
\Photon(255,50)(365,50){3}{7} \Text(310,50)[]{\LARGE $\times$}
\ArrowLine(230,10)(255,50) \ArrowLine(390,10)(365,50)
\Text(240,50)[]{$P_L$} \Text(382,50)[]{$P_L$}
\Text(315,60)[]{$m_{3/2}$} \Text(282.5,37)[]{$\tilde{E}^c$}
\Text(337.5,37)[]{$\tilde{E}$} \Text(255,75)[]{$\tilde{e}^{c*}_1$}
\Text(255,25)[]{$\nu^c_1$} \Text(370,75)[]{$\nu^{c*}_1$}
\Text(370,25)[]{$e^c_1$}
\Text(90,0)[]{\bf{(a)}}  \Text(310,0)[]{\bf{(b)}}
\end{picture}
\caption{Dominant diagram of the bino decay (a) and the gauge
interaction between electrically charged superheavy LR gauginos
and the MSSM lepton singlets (b).}\label{fig:gauginoMed}
\end{center}
\end{figure}
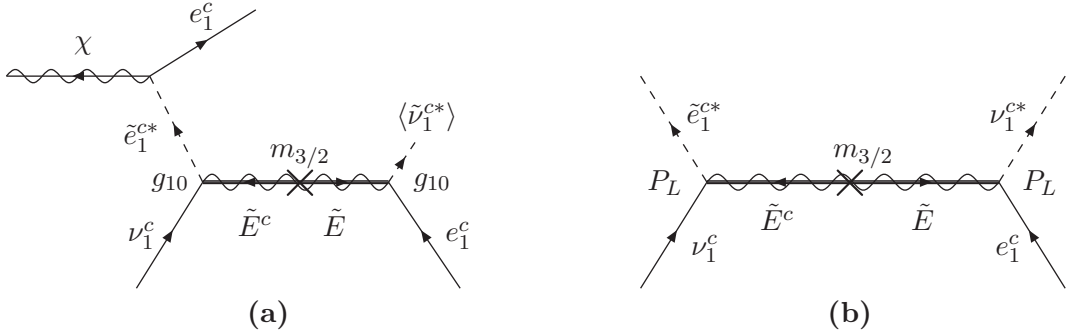

If (1) R-parity is absolutely preserved and (2) $\chi$ is really
the LSP, $\chi$ can never decay. We mildly relax these two
conditions: by assuming a non-zero VEV of the superpartner of the
(first family of) RH neutrino, $\tilde{\nu}^c_1$ (i.e. R-parity
violation), or its mass lighter than the $\chi$'s mass, $m_\chi$
(i.e. $\tilde{\nu}^c_1$ LSP), $\chi$ can decay. By introducing a
global symmetry, one can forbid its renormalizable Yukawa
couplings to the MSSM fields. Then, $\tilde{\nu}^c_1$ can interact
with the MSSM fields only through the superheavy gauge fields and
gauginos of SO(10), since the (s)RH neutrino $\nu^c_1$
($\tilde{\nu}^c_1$) is a {\it neutral singlet} under the SM gauge
symmetry. Consequently, the decay of $\chi$ would be possible but
quite suppressed. For instance, refer to the diagram of
FIG.\ref{fig:gauginoMed}-(a). We will discuss how this diagram can
be dominant for the $\chi$ decay.
%


\begin{table}[!h]
\begin{center}
\begin{tabular}{c} \hline
Interactions of the MSSM fields and heavy gauginos \\
\hline\hline
~$\tilde{e}^{c*}_i\nu^c_i\tilde{E}^c$~, ~~
$\tilde{d}^{c*}_iu^c_i\tilde{E}^c$~,~~
$h_u^{+*}\tilde{h}_d^0\tilde{E}^c$~,~~
$h_u^{0*}\tilde{h}_d^-\tilde{E}^c$~
\\
~~$\tilde{\nu}^{c*}_ie^c_i\tilde{E}$~,~~
$\tilde{u}^{c*}_id^c_i\tilde{E}$~,~ ~
$h_d^{0*}\tilde{h}_u^+\tilde{E}$~, ~~
$h_d^{-*}\tilde{h}_u^0\tilde{E}$~
\\ \hline
$\tilde{\nu}^{c*}_i\nu^c_i\tilde{N}$~,
~$\tilde{u}^{c*}_iu^c_i\tilde{N}$~,
~~$h^{+*}_u\tilde{h}^+_u\tilde{N}$~,
~~$h^{0*}_u\tilde{h}^0_u\tilde{N}$
\\
~$\tilde{e}^{c*}_ie^c_i\tilde{N}$ ,\quad
$\tilde{d}^{c*}_id^c_i\tilde{N}$~,
~~$h^{-*}_d\tilde{h}^-_d\tilde{N}$~,
~~$h^{0*}_d\tilde{h}^0_d\tilde{N}$~
\\ \hline\hline
$\tilde{e}^{c*}_iq_i\tilde{Q}^{'c}$~, ~~
$\tilde{d}^{c*}_il_i\tilde{Q}^{'c}$
~,~~$\tilde{q}^{*}_iu^c_i\tilde{Q}^{'c}$~
\\
$\tilde{q}^*_ie^c_i\tilde{Q}'$~, ~~
$\tilde{l}^*_id^c_i\tilde{Q}'$~,~~
$\tilde{u}^{c*}_iq_i\tilde{Q}'$~
\\
\hline
$\tilde{\nu}^{c*}_iq_i\tilde{Q}^{c}$~, ~~
$\tilde{u}^{c*}_il_i\tilde{Q}^{c}$~, ~~
$\tilde{q}^{*}_id^c_i\tilde{Q}^{c}$ ~
\\
$\tilde{q}^*_i\nu^c_i\tilde{Q}$~, ~~
$\tilde{l}^{*}_iu^c_i\tilde{Q}$~, ~~
$\tilde{d}^{c*}_iq_i\tilde{Q}$~
\\
\hline
$\tilde{u}^{c*}_i\nu^c_i\tilde{U}^c$~, ~~
$\tilde{l}^{*}_iq_i\tilde{U}^c$~, ~~
$\tilde{d}^{c*}_ie^c_i\tilde{U}^c$~
\\
$\tilde{\nu}^{c*}_iu^c_i\tilde{U}$~, ~~
$\tilde{q}^{*}_il_i\tilde{U}$~, ~~
$\tilde{e}^{c*}_id^c_i\tilde{U}$~
\\
\hline
\end{tabular}
\end{center}\caption{Gauge interactions between two MSSM fields and a heavy
gaugino in the SO(10) GUT}\label{tab:heavyInt}
\end{table}


\subsection{The conditions for leptonic decay of $\chi$}

Let us consider the interactions of the superheavy gauginos first.
In TABLE \ref{tab:heavyInt}, we list all the gauge interactions
between the superheavy gauginos of SO(10) and two MSSM fields.
They are, of course, the renormalizable operators. Since
$\tilde{\nu}^c_i$ ($i=1,2,3$) do not couple to $\tilde{Q}^{'c}$
and $\tilde{Q}'$, the interactions by $\tilde{Q}^{'c}$ and
$\tilde{Q}'$ are not directly involved in the $\chi$ decay. As
seen in TABLE \ref{tab:heavyInt}, $\tilde{\nu}^c_i$ or $\nu^c_i$
couples to the superheavy SO(10) gauginos,
$\{\tilde{E},\tilde{E}^c\}$, $\tilde{N}$,
$\{\tilde{Q},\tilde{Q}^c\}$, and $\{\tilde{U},\tilde{U}^c\}$.

According to PAMELA data \cite{PAMELA}, the branching ratio of the
hadronic DM decay modes should not exceed 10 $\%$. To make the
leptonic interactions, i.e. $\tilde{e}^{c*}_i\nu^c_i\tilde{E}^c$,
$\tilde{\nu}^{c*}_ie^c_i\tilde{E}$, and
$\tilde{\nu}^{c*}_i\nu^c_i\tilde{N}$,
$\tilde{e}^{c*}_ie^c_i\tilde{N}$ dominant over the other
interactions in TABLE \ref{tab:heavyInt}, we assume that

$\bullet$ The LR (or ${\rm B-L}$) breaking scale should be lower
than the SU(5) breaking scale, i.e. $\langle {\bf 16}_H\rangle \ll
\langle {\bf 45}_H\rangle$. Then $M_{Q'}$, $M_{Q}$, $M_{U}$ (and
also the masses of the superheavy triplet higgsinos contained in
${\bf 10}_h$) become much heavier than $M_E$ and $M_N$, and so
most of hadronic decay modes of $\chi$ can be easily suppressed
except those by $\tilde{E}^c$, $\tilde{E}$, and $\tilde{N}$ in
TABLE \ref{tab:heavyInt}.

$\bullet$ The slepton $\tilde{e}^c_1$, which composes an SU(2)$_R$
doublet together with $\nu^c_1$, needs to be lighter than the
squarks. Then the decay channels of $\chi$ by
$\tilde{d}^{c*}_iu^c_i\tilde{E}^c$,
$\tilde{u}^{c*}_id^c_i\tilde{E}$, and
$\tilde{u}^{c*}_iu^c_i\tilde{N}$, $\tilde{d}^{c*}_id^c_i\tilde{N}$
become suppressed.  We also require that $\chi$ and
$\tilde{e}^c_1$ are much lighter than the charged MSSM Higgs. So
the leptonic interactions, $\tilde{e}^{c*}_1\nu^c_1\tilde{E}^c$,
$\tilde{\nu}^{c*}_1e^c_1\tilde{E}$, and
$\tilde{\nu}^{c*}_1\nu^c_1\tilde{N}$,
$\tilde{e}^{c*}_1e^c_1\tilde{N}$ can dominate over the others.

$\bullet$ At least one RH neutrino, i.e. the SU(2)$_L$ singlet
neutrino $\nu^c_1$ (and its superpartner $\tilde{\nu}^c_1$) must
be lighter than $\chi$ so that $\chi$ decays to charged leptons.
It is because $\nu^c_i$ is always accompanied by $\tilde{\nu}_i^c$
in the effective operators leading to the leptonic decay of
$\chi$, composed of $\tilde{e}^{c*}_1\nu^c_1\tilde{E}^c$,
$\tilde{\nu}^{c*}_1e^c_1\tilde{E}$, and
$\tilde{\nu}^{c*}_1\nu^c_1\tilde{N}$,
$\tilde{e}^{c*}_1e^c_1\tilde{N}$. If all the sneutrino masses are
heavier than $\chi$, $\tilde{\nu}^c_1$ must develop a VEV for
decay of $\chi$. Once $\nu_1^c$ is light enough, $\tilde{\nu}_1^c$
can achieve a VEV much easily.

To be consistent with PAMELA's observations on high energy
galactic positron excess \cite{PAMELA}, the DM mass should be
around 300 -- 400 GeV \cite{decayPAMELA}.  Thus, one can simply
take the following values;

${\bf 1.}$ $\langle {\bf 16}_H\rangle$ (or
$\langle\tilde{\nu}^c_H\rangle$) $\ll \langle {\bf 45}_H\rangle$.
If $m_{\tilde{\nu}^c_i} > m_\chi$, then $\langle
\tilde{\nu}^c_1\rangle\neq 0$.

${\bf 2.}$ squarks, charged Higgs, higgsinos and other typical
soft masses are of ${\cal O }(1)$ TeV.

${\bf 3.}$ $m_{\nu^c_1}$ $~\ll~$ $m_\chi ~\sim ~$ 300 -- 400 GeV
$~\lesssim~$ $m_{\tilde{e}^c_1}$ $~\ll~$ ${\cal O}(1)$ TeV.

\noindent Consequently, SO(10) is broken first to LR, which would
be the effective gauge symmetry valid below the GUT scale. As seen
from TABLE \ref{tab:heavyInt}, the gauge interactions by the LR
gauginos (and also gauge fields) preserve the baryon numbers. Even
if the masses of the LR gauginos and gauge fields are relatively
light, their gauge interactions don't give rise to proton decay.
We will show later that the decay channels of $\chi$ through the
mediation of the superheavy gauge fields are relatively
suppressed.

\subsection{Seesaw mechanism}

Although one RH neutrino is light enough, the seesaw mechanism for
obtaining the three extremely light physical neutrinos still may
work. Let us consider the following superpotential;
\begin{eqnarray} \label{lep}
W_{\nu}=
%
y^{(\nu)}_{ij}~l_ih_u\nu^c_j(j\neq 1)
+\frac{1}{2}M_{ij}~\nu^c_i\nu^c_j(i,j\neq 1) ,
%
%
\end{eqnarray}
where the Majorana mass term of $\nu^c_i$ could be generated from
the non-renormalizable superpotential $\langle {\bf
\overline{16}}_H\rangle\langle {\bf \overline{16}}_H\rangle {\bf
16}_i{\bf 16}_j/M_P$ ($i,j\neq 1$). Thus, $M_{ij}$ ($\gg\langle
h_u\rangle$) could be determined, if the LR breaking scale by
$\langle {\bf \overline{16}}_H\rangle$ is known. In this
superpotential, we note that one RH neutrino $\nu^c_1$ does not
couple to the MSSM lepton doublets and Higgs. For instance, by
assigning an exotic U(1) R-charge to $\nu_1^c$, one can forbid its
Yukawa couplings to the MSSM superfields. Thus, $\nu^c_1$ would be
decoupled from the other MSSM fields, were it not for the heavy
gauge fields and gauginos of the SO(10) SUSY GUT.

Taking into account only Eq.~(\ref{lep}), one neutrino remains
massless. The two heavy Majorana mass terms of $\nu^c_2$ and
$\nu^c_3$ are sufficient for the other two neutrinos to achieve
extremely small physical masses through the constrained seesaw
mechanism \cite{2RHNseesaw}:
\begin{eqnarray} \label{seesaw}
m_\nu=m_\nu^{T}=-\left(
\begin{array}{ccc}
0 & v_{12} & v_{13}\\
0 & v_{22} & v_{23}\\
0 & v_{32} & v_{33}\\
\end{array} \right)
\left(
\begin{array}{ccc}
0 & 0 & 0\\
0 & M^{-1}_{22} & M^{-1}_{23}\\
0 & M^{-1}_{23} & M^{-1}_{33}\\
\end{array} \right)
\left(
\begin{array}{ccc}
0 & 0 & 0\\
v_{12} & v_{22} & v_{32}\\
v_{13} & v_{23} & v_{33}\\
\end{array} \right) ,
%
%
\end{eqnarray}
where $v_{ij}\equiv y^{(\nu)}_{ij}\langle h_u\rangle$, and
$M^{-1}_{ij}$ denotes the inverse matrix of $M_{ij}$. One of the
eigenvalues of $m_\nu$ is zero and the other two are of order
$v^2/M$. This mechanism provides mixings of order $v/M$ between
the three left-handed and two RH neutrinos.
%
%
Through the diagonalization of the mass matrix in
Eq.~(\ref{seesaw}), the three left-handed neutrinos from the
lepton doublet $l_1$, $l_2$, and $l_3$ can be maximally mixed,
whereas the mixing of the RH neutrinos is only between $\nu^c_2$
and $\nu_3^c$.  A complex phase in $y^{(\nu)}_{ij}$ could make
leptogenesis possible \cite{2RHNseesaw}.

%
%

\subsection{Heavy gauginos' masses}

The gauge interactions between the gauginos and an SU(2)$_R$
lepton doublet (${\bf 2}_{1}$) in the LR model is described by
\begin{eqnarray} \label{LRmodel}
{\cal L}\supset -\frac{1}{2}\left(\tilde{e}^{c*}
~\tilde{\nu}^{c*}\right) \left[
\begin{array}{cc}
g\tilde{N}_{R}+g'\tilde{N}_{BL} & \sqrt{2}g\tilde{E}\\
\sqrt{2}g\tilde{E}^c & -g\tilde{N}_R+g'\tilde{N}_{BL}
\end{array}
\right] \left(
\begin{array}{c}
e^c \\
\nu^c
\end{array}
\right) ~ + ~ {\rm h.c.} ,
\end{eqnarray}
where $\{\tilde{N}_R, \tilde{E}, \tilde{E}^c\}$ and
$\tilde{N}_{BL}$ are the superpartners of the SU(2)$_R$ and
U(1)$_{B-L}$ gauge fields, respectively.
$(-g\tilde{N}_R+g'\tilde{N}_{BL})/\sqrt{g^2+g^{'2}}$ is identified
with ``$\tilde{N}$'' discussed above.  Hence, its orthogonal
component $(g'\tilde{N}_R+g\tilde{N}_{BL})/\sqrt{g^2+g^{'2}}$
corresponds to the bino of the MSSM. The hypercharge of the MSSM
is defined by
\begin{eqnarray}
\frac{Y}{2}=\pm\frac{1}{2}\sigma^3 +\frac{B-L}{2} ,
\end{eqnarray}
where $+$ ($-$) for ${\bf 2}$ (${\bf\overline{2}}$). It is
straightforward to write down the interaction between the LR
gauginos and $\overline{\bf 2}_{-1}$.  When the LR model embedded
in the SO(10) GUT, the LR and ${\rm B-L}$ gauge couplings, $g$ and
$g'$ can be expressed in terms of the SO(10) gauge coupling,
\begin{eqnarray}
g=\sqrt{\frac{2}{3}}g'=g_{10} .
\end{eqnarray}
By introducing a pair of SU(2)$_R$ doublet Higgs [or ${\bf 16}_H$
and $\overline{\bf 16}_H$ in SO(10)],
\begin{eqnarray}
{\bf 2}_{1}=\left(
\begin{array}{c}
e^c_H \\
\nu^c_H
\end{array}
\right)~\subset ~{\bf 16}_H ~,~~{\rm and}~~ \overline{\bf
2}_{-1}=\left(
\begin{array}{c}
e_{H} \\
-\nu_{H}
\end{array}
\right) ~\subset ~ \overline{\bf 16}_H ,
\end{eqnarray}
and, for instance, the superpotential
\begin{eqnarray} \label{LRbreak}
W=S\left({\bf 2}_{1}\overline{\bf 2}_{-1}-M_{\rm LR}^2\right)
~\subset ~ S\left({\bf 16}_H\overline{\bf 16}_H -M_{\rm
LR}^2\right) ,
\end{eqnarray}
one can break LR to the MSSM gauge group. Here, $S$ is a singlet
superfield.
By non-vanishing VEVs along the neutrino direction and the
``D-flat'' condition, $\langle\tilde{\nu}^{c}_{H}\rangle
=\langle\tilde{\nu}^c_{H}\rangle =v/\sqrt{2}$, $\{e^c_H,
\tilde{E}\}$ and $\{e_H, \tilde{E}^c\}$ obtain the same Dirac
masses, and also the neutral gaugino $\tilde{N}$ and
$(\nu^c_H-\nu_H)/\sqrt{2}$ ($\equiv \nu^c_-$) achieve a mass:
\begin{eqnarray} \label{mass}
&&\quad\quad\quad -{\cal L}_{mass}=M_E\left( e^c_{H}\tilde{E} +
e_{H}\tilde{E}^c\right)+M_N\tilde{N}\nu^c_-
\\
&&+~ m_{3/2}\tilde{E}\tilde{E}^c +
\frac{1}{2}m_{3/2}'\tilde{N}^2+m_{3/2}''\left(e^c_{H}e_H
+\frac{1}{2}\nu^c_{-}\nu^c_-\right)~+~ {\rm h.c.} , \nonumber
\end{eqnarray}
where $M_E\equiv vg_{10}/2$ and $M_N\equiv
v\sqrt{g^2+g^{'2}}/2=vg_{10}\sqrt{5/8}=M_E\sqrt{5/2}$. We note
here that $M_N$ is heavier than $M_E$. The other combination
$(\nu^c_H+\nu_H)/\sqrt{2}$ ($\equiv \nu^c_+$) and $S$ get a mass
from the superpotential Eq.~(\ref{LRbreak}) at the SUSY minimum.
The second line of Eq.~(\ref{mass}) contains the soft mass terms.
Since $S$ can develop a VEV of order the gravitino mass $m_{3/2}$
due to the ``A-term'' corresponding to $W$ of Eq.~(\ref{LRbreak}),
the last two mass terms of Eq.~(\ref{mass}) [$\subset
\langle\tilde{S}\rangle\left(e^c_{H}e_H -\nu^c_{H}\nu_H\right)$]
are induced. We rewrite Eq.~(\ref{mass}) in terms of the four
component spinors as follows;
\begin{align}
-{\cal L}_{mass}= \left(\overline{\lambda^-_R}
~\overline{\psi^-_{R}}\right) \left[
\begin{array}{cc}
m_{3/2} & M_{E} \\
M_E & m''_{3/2}
\end{array} \right]
\left(
\begin{array}{c}
\lambda^-_L \\
\psi^-_L
\end{array}
\right) +\frac{1}{2} \left(\overline{\lambda^0_R}
~\overline{\psi^0_{R}}\right) \left[
\begin{array}{cc}
m_{3/2}' & M_{N} \\
M_N & m''_{3/2}
\end{array} \right]
\left(
\begin{array}{c}
\lambda^0_L \\
\psi^0_L
\end{array}
\right) ~+~ {\rm h.c.}
\end{align}
where $\lambda^{-(0)}$ and $\psi^{-(0)}$ are the Dirac (Majorana)
spinors constructed with the two components' Weyl spinors for the
gauginos and higgsinos:
\begin{eqnarray}
\lambda^-=\left(
\begin{array}{c}
\tilde{E} \\
\overline{\tilde{E}^c}
\end{array}
\right) ~,~~ \psi^-=\left(
\begin{array}{c}
e_H \\
\overline{e^c_{H}}
\end{array}
\right) ~,~~{\rm and}~~ \lambda^0=\left(
\begin{array}{c}
\tilde{N} \\
\overline{\tilde{N}}
\end{array}
\right) ~,~~ \psi^0=\left(
\begin{array}{c}
\nu^c_- \\
\overline{\nu^c_-}
\end{array}
\right) ,
\end{eqnarray}
where the ``bar'' denotes the complex conjugates of the fermionic
fields. $\lambda^+$ and $\psi^+$ are respectively given by
$(\lambda^-)^C$ and $(\psi^-)^C$, and $\lambda^0$ and $\psi^0$
satisfy $(\lambda^0)^C=\lambda^0$ and $(\psi^0)^C=\psi^0$. The
mass eigenstates and their eigenvalues turn out to be
\begin{eqnarray} \label{Estate}
\left(
\begin{array}{c}
\Lambda_1^{-,0} \\
\Lambda_2^{-,0}
\end{array}
\right)_L = \frac{1}{\sqrt{2}}\left[
\begin{array}{cc}
1-\epsilon ~&~ -(1+\epsilon) \\
1+\epsilon ~&~ 1-\epsilon
\end{array} \right]\left(
\begin{array}{c}
\lambda^{-,0} \\
\psi^{-,0}
\end{array}
\right)_L ,~ {\rm and}~~ \quad\quad~~
\\
M_{1,2}^{(-)}= \mp M_{E}+\frac{1}{2}\left[m_{3/2}+m''_{3/2}\right]
~, ~~ M_{1,2}^{(0)}= \mp
M_{N}+\frac{1}{2}\left[m_{3/2}'+m''_{3/2}\right] , \label{Evalue}
\end{eqnarray}
where $\epsilon\equiv [m_{3/2}^{(')}-m''_{3/2}]/(4M_{E,N})$ ($\ll
1$).
%


\subsection{Heavy gauginos' propagations}

From Eq.~(\ref{LRmodel}), the charged interactions read as
\begin{eqnarray} \label{cc}
-{\cal
L}_{c.c.}=\frac{g}{\sqrt{2}}\left(\tilde{e}^{c*}_i\nu^c_i\tilde{E}^c
+ \tilde{\nu}^{c*}_ie^c_i\tilde{E} + {\rm h.c.} \right) =
\frac{g_{10}}{\sqrt{2}}\left[\tilde{e}^{c*}_i~\overline{\lambda^-}P_L(\nu_{Di})^C
+ \tilde{\nu}^{c*}_i~\overline{e^-_{Di}}P_L\lambda^- + {\rm h.c.}
\right] ,~~
\end{eqnarray}
where $P_L$ stands for the projection operator. $\nu_{Di}$ and
$e^-_{Di}$ are Dirac spinors defined as
\begin{eqnarray}
\nu_{Di}=\left(
\begin{array}{c}
\nu \\
\overline{\nu^c}
\end{array}
\right)_i ~~{\rm and}~~ e^-_{Di}=\left(
\begin{array}{c}
e^- \\
\overline{e^c}
\end{array}
\right)_i .
\end{eqnarray}
By contraction of $\lambda^-$ and $\overline{\lambda^-}$ in
Eq.~(\ref{cc}), therefore, the effective operator leading to
$\tilde{e}^{c*}_1\rightarrow e^-_1 +\nu_1+\tilde{\nu}^c_1$ is
induced. See the diagram of FIG.\ref{fig:gauginoMed}-(b).
$\lambda^-$ is decomposed to the two mass eigenstates
$\Lambda_1^-$ and $\Lambda_2^-$, as shown in Eq.~(\ref{Estate}).
With Eqs.~(\ref{Estate}) and (\ref{Evalue}), the amplitude
suppression coming from the superheavy gaugino's propagator
$\langle T\lambda^-\overline{\lambda^-}\rangle$ is estimated as
\begin{eqnarray}
i\left(\frac{1-\epsilon}{\sqrt{2}}\right)^2P_L
\frac{{\not}p+M_1}{p^2-M_1^{2}}P_L
+i\left(\frac{1+\epsilon}{\sqrt{2}}\right)^2P_L
\frac{{\not}p+M_2}{p^2-M_2^{2}}P_L \quad \approx \quad
i~\frac{m_{3/2}}{M_E^2}P_L
\end{eqnarray}
at low energies.  Thus, the decay, $\tilde{e}^{c*}_1\rightarrow
e^-_1 +\nu_1+\tilde{\nu}^c_1$ is extremely suppressed, but still
possible if it is kinematically allowed.

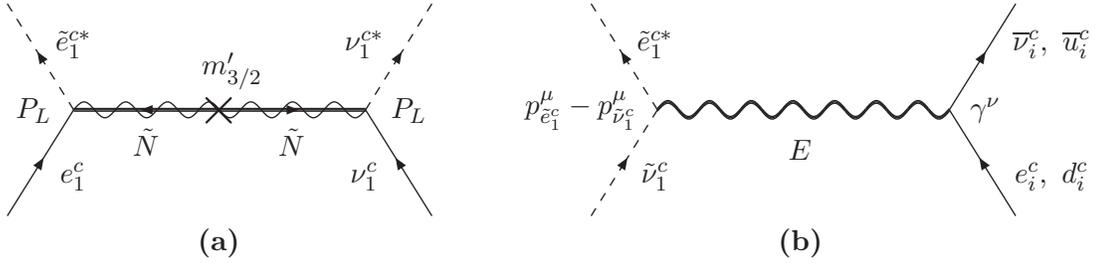
\begin{figure}[t]
\begin{center}
\begin{picture}(410,110)(0,0)

\DashArrowLine(35,50)(10,90){3} \DashArrowLine(145,50)(170,90){3}
%
\ArrowLine(90,50)(35,50) \ArrowLine(90,50)(145,50)
\Line(145,50.5)(35,50.5) \Line(145,49.5)(35,49.5)
\Photon(35,50)(145,50){3}{7} \Text(90,50)[]{\LARGE $\times$}
\ArrowLine(10,10)(35,50) \ArrowLine(170,10)(145,50)
\DashArrowLine(255,50)(230,90){3} \ArrowLine(365,50)(390,90)
%
\Photon(255,50.5)(365,50.5){3}{7}
\Photon(255,49.5)(365,49.5){3}{7} \Photon(255,50)(365,50){3}{7}
\DashArrowLine(230,10)(255,50){3} \ArrowLine(390,10)(365,50)
\Text(20,50)[]{$P_L$} \Text(162,50)[]{$P_L$}
\Text(95,65)[]{$m'_{3/2}$} \Text(62.5,38)[]{$\tilde{N}$}
\Text(117.5,38)[]{$\tilde{N}$} \Text(35,75)[]{$\tilde{e}^{c*}_1$}
\Text(35,25)[]{$e^c_1$} \Text(145,75)[]{$\nu^{c*}_1$}
\Text(145,25)[]{$\nu^c_1$}
\Text(227,50)[]{$p^\mu_{\tilde{e}_1^c}-p^\mu_{\tilde{\nu}_1^c}$}
\Text(380,50)[]{$\gamma^\nu$} \Text(310,35)[]{$E$}
\Text(255,75)[]{$\tilde{e}^{c*}_1$}
\Text(255,25)[]{$\tilde{\nu}^c_1$}
\Text(405,75)[]{$\overline{\nu}^c_i,~\overline{u}^c_i$}
\Text(405,25)[]{$e^c_i,~d^c_i$}

\Text(90,0)[]{\bf{(a)}}  \Text(310,0)[]{\bf{(b)}}

\end{picture}
\caption{The gauge interaction between the electrically neutral
superheavy gaugino and the MSSM lepton singlets (a) and between
gauge fields and the SU(2)$_L$ singlets of the MSSM.
}\label{fig:gaueMed}
\end{center}
\end{figure}

Eq.~(\ref{LRmodel}) includes also the neutral interactions of the
SU(2)$_L$ lepton singlets with $\tilde{N}$ and the bino. One can
extract the part interacting only with $\tilde{N}$:
\begin{eqnarray} \label{nc}
-{\cal L}_{n.c.}=
\frac{\sqrt{g^2+g^{'2}}}{2}\tilde{\nu}^{c*}_i\nu^c_i\tilde{N} +
\frac{g^2-g^{'2}}{2\sqrt{g^2+g^{'2}}}\tilde{e}^{c*}_ie^c_i\tilde{N}
 + {\rm h.c.} \nonumber \\
=\frac{g_{10}}{\sqrt{2}}\left[\frac{\sqrt{5}}{2}\tilde{\nu}^{c*}_i
~\overline{\lambda^0}P_L(\nu_{Di})^C
+\frac{1}{\sqrt{20}}\tilde{e}^{c*}_i~\overline{e^-_{Di}}P_L\lambda^0
+ {\rm h.c.} \right] .
\end{eqnarray}
They are actually reminiscent of the $Z$ boson interactions in the
SM. By contracting $\lambda^0$ and $\overline{\lambda^0}$, the
decay $\tilde{e}^{c*}_1\rightarrow e^-_1 +\nu_1+\tilde{\nu}^c_1$
is possible. See FIG.\ref{fig:gaueMed}-(a). However, since
$M_{N}^2$ is $\frac{5}{2}$ times heavier than $M_E^2$ as shown
from Eq.~(\ref{mass}), and the effective coupling is
$\frac{\sqrt{5}}{2}\times\frac{1}{\sqrt{20}}=\frac{1}{4}$ times
smaller than that of the charged interaction case, the amplitude
mediated by $\lambda^0$ is just $\frac{1}{10}$ of that by
$\lambda^-$.


As seen in TABLE \ref{tab:heavyInt}, the MSSM Higgs and higgsinos
also couple to $\tilde{E}^c$, $\tilde{E}$ or $\tilde{N}$. Since
the MSSM charged Higgs and higgsinos are assumed to be much
heavier than $\tilde{e}^c_1$ and $\chi$, the decay channels
through them are quite suppressed or kinematically forbidden.
%

So far we did not discuss the case in which $\chi$ decays through
the mediation of the superheavy gauge bosons. The potentially
dominant diagram is displayed in FIG.\ref{fig:gaueMed}-(b).
$\tilde{e}^c_1$ is coupled to $\chi$ and $e^c_1$. The
scalar-scalar-gauge boson vertex is basically a derivative
coupling. Accordingly, this diagram is suppressed compared to
FIG.\ref{fig:gauginoMed}-(b), only if the bino is much lighter
than the soft mass of $\{\tilde{E},\tilde{E}^c\}$. As presented
above, in this paper, we assume that $m_\chi\sim 300$ -- 400 GeV
and the soft mass of $\{\tilde{E},\tilde{E}^c\}$ is of ${\cal
O}(1)$ TeV.

\subsection{LSP decay rate and the seesaw scale}

Now let us estimate the decay rate of
FIG.\ref{fig:gauginoMed}-(a), which is the dominant decay channel,
and determine the LR breaking scale such that it is consistent
with PAMELA data. Indeed, if $m_{\tilde{\nu}^c_1} < m_\chi$, a
non-zero VEV of $\tilde{\nu}^c_1$ is not essential: $\chi$ can
decay to the four light particles, $e^\pm$, $\nu^c_1$, and
$\tilde{\nu}^c_1$. However, just for simplicity, we will assume
that a non-zero VEV of $\tilde{\nu}^c_1$ is developed.  For
instance, let us consider the following terms in the
superpotential;
\begin{eqnarray} \label{nuVEV}
W\supset \frac{1}{M_P}\langle\overline{\bf 16}_H\rangle{\bf
16}_1\Sigma^2+\kappa\Sigma^3,
\end{eqnarray}
where $M_P=2.4\times 10^{18}$ GeV and $\kappa$ is a dimensionless
coupling constant. $\Sigma$ is an SO(10) singlet. We assign e.g.
the U(1) R-charge of $2/3$ to ${\bf 16}_1$ and $\Sigma$, and $0$
to $\overline{\bf 16}_H$.
%
%
The scale of $\langle\overline{\bf 16}_H\rangle$ ($=\langle {\bf
16}_H\rangle = M_E/\sqrt{2}g_{10}$) can be determined such that it
is consistent with PAMELA data. The soft mass term of $\Sigma$ and
the A-term corresponding to $\kappa\Sigma^3$ in the scalar
potential permit a VEV $\langle\tilde{\Sigma}\rangle\sim
m_{3/2}/\kappa$. Then, the scalar potential generates a linear
term of $\tilde{\nu}^c_1$ coming from the A-term corresponding to
the first term of Eq.~(\ref{nuVEV}), $V\supset m_{3/2}^3(\langle
\overline{\bf 16}_H\rangle/\kappa^2M_P) \tilde{\nu}^c_1$. The
linear term and the soft mass term of $\tilde{\nu}^c_1$ in the
scalar potential can induce a non-zero VEV of $\tilde{\nu}^c_1$:
\begin{eqnarray}
\langle\tilde{\nu}^c_1\rangle \sim \frac{m_{3/2}}{\kappa^2}\times
\frac{M_E}{M_P} .
\end{eqnarray}
Thus, the decay rate of $\chi$ in FIG.\ref{fig:gauginoMed}-(a) can
be estimated:
\begin{eqnarray}
\Gamma_\chi=\frac{\alpha_{10}^2\alpha_Ym_\chi^5}{96M_E^4}
\left(\frac{m_{3/2}\langle
\tilde{\nu}^c_1\rangle}{m_{\tilde{e}^c_1}^2}\right)^2\sim
\frac{\alpha_{10}^2\alpha_Ym_\chi^5}{96M_E^2M_P^2}
\left(\frac{m_{3/2}}{\kappa m_{\tilde{e}^c_1}}\right)^4 \sim
10^{-26}~{\rm sec.}^{-1},
\end{eqnarray}
where $\alpha_{10}$ ($\equiv g_{10}^2/4\pi$) and $\alpha_Y$
[$\equiv g_Y^2/4\pi=(3/5)\times g_1^2/4\pi$, where $g_1$ is the
SO(10) normalized gauge coupling of $g_Y$] are approximately
$1/24$ and $1/100$, respectively. Here, we ignore the RG
correction to $\alpha_{10}$.
300 -- 400 GeV fermionic DM  decaying to $e^\pm$ and a light
neutral particle can fit the PAMELA data \cite{decayPAMELA}. For
$m_\chi\approx $ 300 -- 400 GeV, $(m_{3/2}/\kappa
m_{\tilde{e}^c_1}) \sim 10$, $M_E$ or $\langle {\bf 16}_H\rangle$
is estimated to be of order $10^{14}$ GeV. This is consistent with
the assumption $\langle{\bf 16}_H\rangle\ll\langle{\bf
45}_H\rangle\sim 10^{16}$ GeV.
Therefore, the masses of the other two RH neutrinos, which do not
contribute to the process of FIG.\ref{fig:gauginoMed}-(a), are
around $10^{10}$ GeV or smaller in this case: $W\supset
y_{ij}(\langle \overline{\bf 16}_H\rangle\langle \overline{\bf
16}_H\rangle/M_P) {\bf 16}_i{\bf 16}_j (i,j\neq 1)\supset
y_{ij}(10^{10}~{\rm GeV})\times \nu^c_i\nu^c_j (i,j\neq 1)$. So
the Yukawa couplings of the Dirac neutrinos should be a bit small
($\sim 10^{-2}$).

If $m_\chi\approx 3.5$ TeV and the model is slightly modified such
that $\chi$ decays dominantly to $\mu^\pm,~\nu^c_2$ rather than to
$e^\pm,~\nu^c_1$, which is straightforward, the Fermi-LAT's data
as well as the PAMELA's can be also explained \cite{decayPAMELA}.
In this case, $M_E$ or $\langle {\bf 16}_H\rangle$ should become
somewhat heavier ($\sim 10^{15}$ GeV), and the seesaw scale should
be replaced by $10^{12}$ GeV. However, the motivation of
introducing SUSY to resolve the gauge hierarchy problem in the SM
would become more or less spoiled.

%
%
%

\section{Conclusions}

In this paper, we have shown that the bino-like LSP in the MSSM
can decay through the SO(10) gauge interactions, if a RH neutrino
is light enough ($m_{\nu^c_1}\lesssim m_\chi$) and its
superpartner develops a VEV ($\langle \tilde{\nu}^c_1\rangle\neq
0$).
The Yukawa couplings between the RH (s)neutrino and the MSSM
fields can be suppressed by a global symmetry such as the U(1)
R-symmetry. It gives rise to an extremely small R-parity violation
very naturally.
If the LR breaking scale or the seesaw scale is low enough
compared to the GUT scale (i.e. $\langle {\bf 16}_H\rangle\ll
\langle {\bf 45}_H\rangle\sim 10^{16}$ GeV), and squarks, the MSSM
charged Higgs, higgsinos, and other typical soft masses are
relatively heavier ($\sim {\cal O}(1) $ TeV) than the slepton, the
recently reported PAMELA's high energy galactic positrons can be
explained through the leptonic decay of the bino-like LSP in the
framework of the SO(10) SUSY GUT. Particularly, we assumed the
quite mild hierarchies for the (s)lepton mass parameters;
$m_{\nu^c_1}\ll m_\chi\sim$ 300 -- 400 GeV $\lesssim
m_{\tilde{e}^c_1}\ll {\cal O}(1)$ TeV.
In the bench mark model, $\langle {\bf 16}_H\rangle\sim {\cal O
}(10^{14})$ GeV, and the two RH neutrino masses turned out to be
of order $10^{10}$ GeV or smaller. Even if one RH (s)neutrino is
almost decoupled from the interactions of the MSSM, the extremely
light three physical neutrinos and their oscillations still can be
achieved through the seesaw mechanism.

\acknowledgments{ \noindent The author is supported by the FPRD of
the BK21 program, and in part by the Korea Research Foundation,
Grant No. KRF-2005-084-C00001. }


\end{document}